\documentstyle[aps,prb,multicol,psfig,array]{revtex}
\tighten
\begin{document}
\title{First-order transition between a small-gap semiconductor and a
ferromagnetic metal in the isoelectronic alloys FeSi$_{1-x}$Ge$_x$}

\author{V. I. Anisimov$^{1,2}$, R. Hlubina$^{1,3}$, M. A. Korotin$^{2}$,
V. V. Mazurenko$^{1,2}$, T. M. Rice$^1$, A.O.Shorikov$^{1,2}$, 
M. Sigrist$^1$}

\address{$^1$Institute for Theoretical Physics,
ETH H\"onggerberg, CH-8093 Z\"urich, Switzerland}

\address{$^2$Institute of Metal Physics, Russian Academy of Sciences,
620219 Yekaterinburg GSP-170, Russia}

\address{$^3$Department of Solid State Physics, Comenius University,
Mlynsk\'{a} Dolina F2, 842 48 Bratislava, Slovakia}

\maketitle

\begin{abstract}
The contrasting groundstates of isoelectronic and isostructural FeSi
and FeGe can be explained within an extended local density
approximation scheme (LDA+U) by an appropriate choice of the onsite
Coulomb repulsion, $U$ on the Fe-sites. A minimal two-band model with
interband interactions allows us to obtain a phase diagram for the
alloys FeSi$_{1-x}$Ge$_{x}$.  Treating the model in a mean field
approximation, gives a first order transition between a small-gap
semiconductor and a ferromagnetic metal as a function of magnetic
field, temperature, and concentration, $x$. Unusually the transition
from metal to insulator is driven by broadening, not narrowing, the
bands and it is the metallic state that shows magnetic order.
\end{abstract}
\pacs{PACS}
\begin{multicols}{2}
The unusual magnetic susceptibility of FeSi has been a subject of
interest for a long time. Jaccarino {\it et al.}~\cite{Jaccarino67}
found a rapid crossover around room temperature from activated
behavior at low temperature to an apparent localized Curie-Weiss form
at high temperature. This has stimulated a number of proposed
explanations. Among them are a form of Kondo insulating behavior in
FeSi~\cite{Mason92,Mandrus95}, although an underlying microscopic
model has not been clarified, and almost ferromagnetic semiconductor
model~\cite{Takahashi79,Evangelou83}. The latter received support from
the detailed LDA band structure calculation of Mattheiss and
Hamann~\cite{Mattheiss93}. They found a small gap semiconductor with
formal neutral valences Fe$^{0+}$, Si$^{0+}$ on both elements.  Later
Anisimov {\it et al.}~\cite{Anisimov96} extended these calculations to
include the onsite Coulomb interaction, $U$, in a mean field
approximation, the so-called LDA+U method~\cite{lda+u}, and found, for
a reasonable choice of $U$, a ferromagnetic metallic phase very close
by in energy. This led them to propose that a transition to this
ferromagnetic state could be driven by applying a magnetic field ($B$)
and that the unusual temperature ($T$) dependent susceptibility in
FeSi reflected the proximity to a critical point of this transition at
finite ($B_c$, $T_c$). However the predicted value of $B_c$
($\approx$170T) is too large to be reached in the laboratory and the
existence of the critical point remains untested experimentally. A
non-trivial prediction was the fractional value of the saturation
moment of the ferromagnetic phase at $S=1/2$ or 1 $\mu_B$/Fe. This is
a consequence of the band structure and cannot be reconciled with a
local model and a 3d$^8$ configuration for Fe$^{0+}$.

The magnetic properties of the isoelectronic and isostructural FeGe
compound have aroused less interest. It is a magnetic metal with a
long period spiral form which is simply a ferromagnet twisted by the
Dzyaloshinskii-Moriya interaction that is a consequence of the absence
of inversion symmetry at the Fe site in this cubic
structure~\cite{Lebech89}. Interestingly the saturation moment is the
fractional $S=1/2$ value quoted above
~\cite{Wappling68,Lundgren68}. In this letter we report LDA+U
calculations (in TBLMTO calculation scheme \cite{lmto}) for FeGe and
the isoelectronic alloys FeSi$_{1-x}$Ge$_{x}$. Motivated by these
results we use a simplified phenomenological model to explore the
complete phase diagram of the isoelectronic alloys FeSi$_{1-x}$Ge$_x$
with varying $T$ and $B$.  Alloying FeSi with Ge allows one to tune the
predicted semiconductor to ferromagnetic metal transition and its
unusual critical point at ($B_c$, $T_c$), to experimentally convenient
values.

\vspace{ 0.5cm }
\begin{figure}
\centerline{\psfig{figure=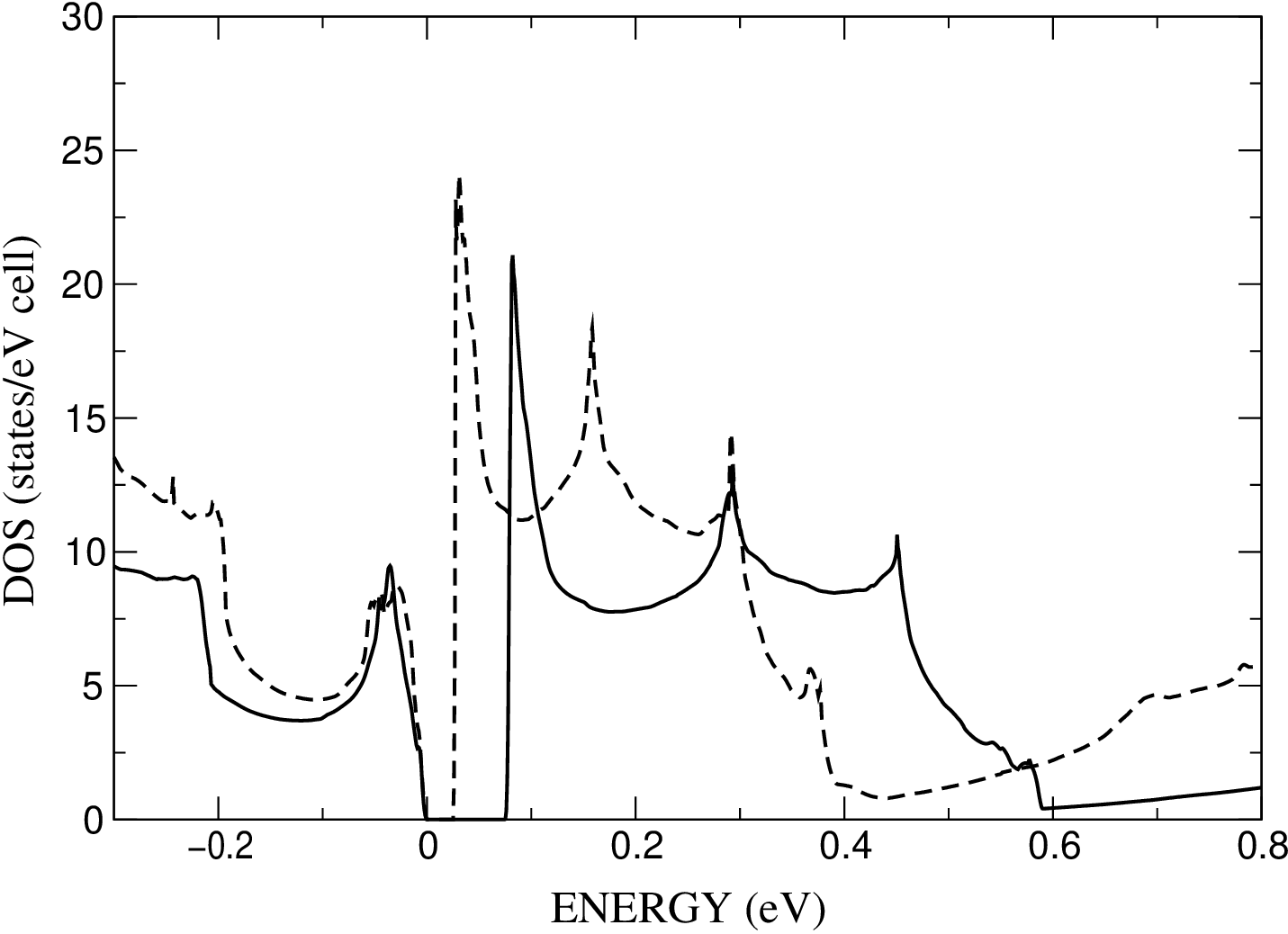,width=0.4\textwidth,angle=-90}}
{\small FIG. 1: The density of states (DOS) near the Fermi level at
zero energy obtained from LDA calculations.  The solid and dashed
lines represent FeSi and FeGe with energy gaps 0.08 eV and 0.03 eV
respectively.  The corresponding widths of the peaks above the Fermi
level are 0.5 eV and 0.37 eV .}
\end{figure}

Both FeSi and FeGe crystallize in the cubic B20 structure which can be
viewed as highly distorted rocksalt with 4 FeSi formula units in the
primitive cell. Each Fe site has 7 Si neighbors and point group
symmetry, C$_3$. The LDA band structure calculations of Mattheiss and
Hamann \cite{Mattheiss93} place the Fermi energy within a narrow
manifold of 20 bands of predominantly Fe 3$d$ character lying within a
larger hybridization gap with the Si ($3p, 3s$) states.  A fuller
description of the band structure will be presented elsewhere
\cite{future}. A nontrivial feature of their results is the presence
of a small but complete band gap separating 16 filled valence bands
from 4 empty conduction bands. The conduction bands have nonbonding
character with respect to the Si atoms and are divided by a pseudogap
into two lower lying quite narrow bands and two higher lying bands
which are much wider. The results of our LDA band calculations for
FeGe are similar but with an even smaller band gap and narrower
overall bandwidths as expected from the larger lattice constant. The
total density of states for both compounds is shown in Fig.~1.

Anisimov et al.~\cite{Anisimov96} extended the LDA calculations for
FeSi to incorporate the onsite Coulomb repulsion ($U$) on the Fe-sites
in a mean field approximation scheme known as LDA+U \cite{lda+u}. They
found a local minimum in the total energy versus magnetization at a
value $\approx 1 \mu_B$/Fe~ which moved to lower energy with
increasing $U$.  In this magnetized state the lower pair of conduction
bands are filled for the majority spin direction while the Fermi level
of the minority spin electrons lies in the valence band complex
leading to metallic behavior. The relative positions of the two minima
corresponding to the small gap semiconductor and magnetized metal,
depends on the choice of $U$ which is not known precisely.
\vspace{0.5 cm }
\begin{figure}
\centerline{\psfig{figure=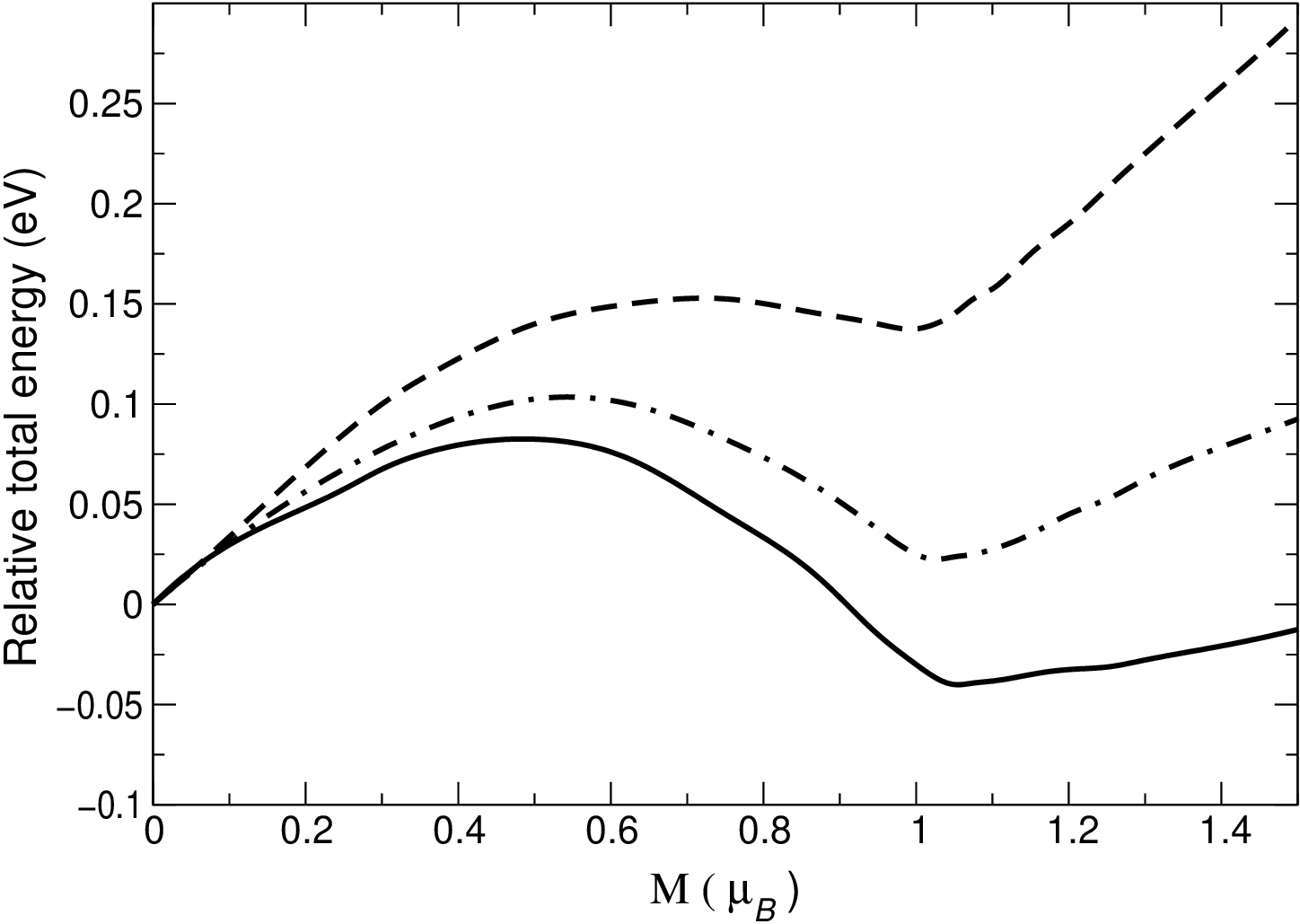,width=0.4\textwidth,angle=-90}}
{\small FIG. 2: The evolution of the total energy ( with energy of
nonmagnetic solution taken a zero ) as a function of the spin moment
$M (\mu_{B}/Fe)$ for the value of $U=3.7$eV. The solid, dashed and
dashed-dotted lines correspond to FeGe, FeSi and
FeSi$_{0.58}$Ge$_{0.42}$ respectively.}
\end{figure}

We have used the TBLMTO scheme \cite{lmto} to perform LDA+U
calculations for FeGe.  The narrower bandwidths and energy gap lead to
a lower value for the critical $U$, where the two energy minima cross,
in FeGe relative to FeSi. Thus it is possible to obtain within the
LDA+U scheme for a common value of $U$, the correct groundstates for
FeSi and FeGe \cite{DM} (see Fig.~2). There are two unusual aspects to
this metal-insulator transition (MIT). It is the metallic state, rather than
the insulating state, which is magnetically ordered. Secondly, the
transition from insulating to metallic behavior is driven by
narrowing, rather than increasing, the bandwidth. The inverted nature
of the metal-insulator transition is a direct consequence of the fact
that the MIT is driven by the paramagnet-ferromagnet transition.

FeSi and FeGe are the end members of the isoelectronic and
isostructural alloys, FeSi$_{1-x}$Ge$_x$. We have extended the LDA+U
calculations to the alloys \cite{alloy} using experimental lattice
parameters \cite{exp_alloy}.  The critical value $x_c$ of the first
order transition is sensitive to the choice of $U$. For the value of
$U=3.7$eV, illustrated in Fig.~2, $x_c=0.4$ in a good agreement with
the experimental value $x_c=0.3$ \cite{exp_alloy}.

To proceed further we introduce and solve (within a mean field
approximation) a minimal phenomenological model for the isoelectronic
alloys FeSi$_{1-x}$Ge$_x$.  Motivated by the experimental fact that
the ordered moment in FeGe is $\approx 1\mu_B$ per Fe
atom~\cite{Wappling68,Lundgren68} which implies a fractional
magnetization with respect to the paramagnetic Fe $3d^8$
configuration, we are led to an itinerant model of magnetism.  We
consider the following model for the conduction and valence bands.
The valence band, describing the upper part of the manifold of
occupied Fe $3d$ bands, stretching below zero energy in Fig.~1, has a
width $2W$ and contains one state per spin per Fe atom.  The
conduction band (of width $W$) contains $1/2$-state per spin per Fe
atom and when multiplied by 4 (which is the number of Fe atoms per
primitive cell), it corresponds to the two narrow conduction bands
just above the Fermi level.  Both the valence and conduction bands are
assumed for simplicity to have a constant density of states
$N(\varepsilon)=(2W)^{-1}$ per spin, with a ${\bf k}$-independent gap
of $2\Delta$.  Taking 2 conduction electrons per Fe atom, leads to a
fully occupied valence band and a semiconducting ground state in the
noninteracting case.

\vspace{ 0.5cm }
\begin{figure}
\centerline{\psfig{figure=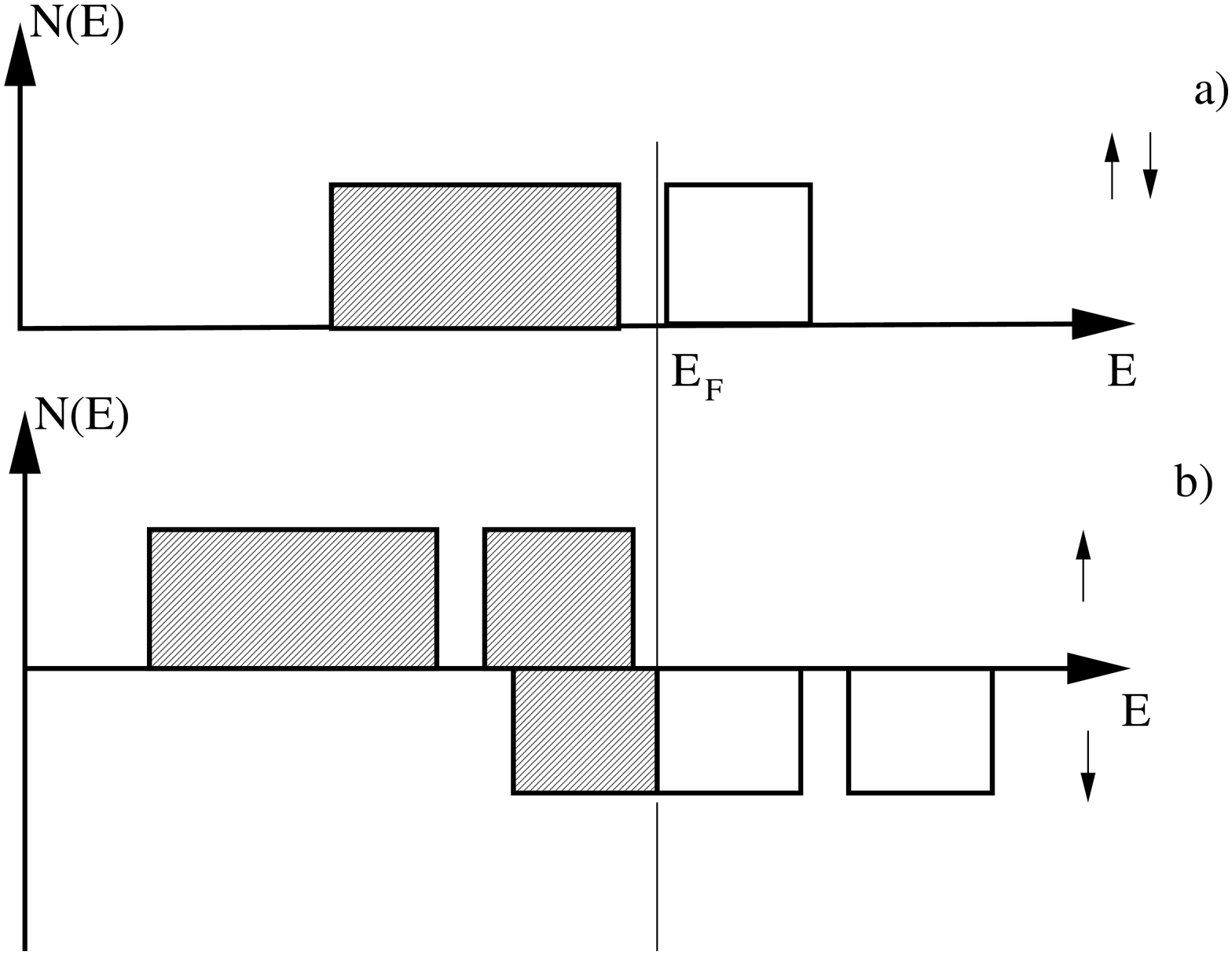,width=0.4\textwidth}} {\small
FIG. 3: Density of states of the model band structure.  (a)
Nonmagnetic semiconducting state. (b) Magnetic metallic state. The
arrows note majority and minority spins.}
\end{figure}

We propose that three types of effective local interactions should
control the physics: intraband repulsion $\widetilde{U}$, interband
repulsion $V$, and exchange coupling between the bands $I$. The
interactions $\widetilde{U}$, $V$, and $I$ are combinations of the
matrix elements of the Coulomb interaction between the relevant
Wannier orbitals and are taken as a phenomenological input to our
theory.  Furthermore, we assume that the primary effect of the
$\widetilde{U}$ term is to strongly renormalize the bandwidth $W$ with
respect to estimates from the LDA calculations. In order to describe
the experimental data, we have therefore decided to use a
phenomenological bandwidth rather than the calculated $W\sim
0.5$~eV~\cite{Mattheiss93}. Thus, in order not to double count its
effects, the $\widetilde{U}$ term is not included in our model
Hamiltonian which reads
\begin{eqnarray}
\nonumber
H=\sum_{\lambda,{\bf k},\sigma}
(\varepsilon_{\lambda {\bf k}}-\sigma\mu_B B)n_{\lambda {\bf k} \sigma}
-{1\over 2}\sum_{\bf R}\left(I\hat{m}_{\bf R}^2+
V\hat{\rho}_{\bf R}^2\right),
\end{eqnarray}
where $\lambda=-1,+1$ denotes the valence and conduction bands,
respectively, and $\varepsilon_{\lambda {\bf k}}$ and $n_{\lambda {\bf
k} \sigma}$ is the band energy and number of electrons with momentum
${\bf k}$ and spin $\sigma$ in band $\lambda$.  The operators
$\hat{\rho}_{\bf R}=2+\sum_{\lambda \sigma}\lambda n_{\lambda {\bf R}
\sigma}$ and $\hat{m}_{\bf R}=\sum_{\lambda \sigma} \sigma n_{\lambda
{\bf R} \sigma}$ measure the local number of charge carriers and the
local magnetization, respectively.  The spin quantization axis has
been chosen in the direction of the external magnetic field $B$. The
Hamiltonian can be written also in an explicitly spin-rotation
invariant form, which we do not specify here.

Below we describe the phase diagram of the FeSi$_{1-x}$Ge$_x$ alloys
by fixing the values of $I$, $V$, and $\Delta$ irrespective of the Ge
content $x$. Alloying changes the bandwidth $W$ which decreases with
increasing $x$, in accordance with the LDA results. Treating the
interaction terms in the mean field approximation allows to replace
the operators $\hat{\rho}_{\bf R}$ and $\hat{m}_{\bf R}$ by the
expectation values $\rho$ and $m$, respectively.  The main technical
difference of the present study with respect to
Ref.~\onlinecite{Anisimov96} is that our model is not particle-hole
symmetric, and therefore in addition to $m$ and $\rho$, the chemical
potential $\mu$ has to be calculated self-consistently.  Assuming
finite values of $m$ and $\rho$, the single-particle energies in the
lower and upper bands are $\omega_{\lambda,\sigma}=\varepsilon-\lambda
V\rho-\sigma(\mu_BB+Im)$.  The corresponding mean particle numbers are
\begin{eqnarray}
\langle n_{-,\sigma}\rangle&=&
(2W)^{-1}\int_{-\Delta-2W}^{-\Delta}{d\varepsilon\over
{1+\exp(\omega_{-,\sigma}-\mu)/T}},
\label{eq:n2}
\\
\langle n_{+,\sigma}\rangle&=&
(2W)^{-1}\int_{\Delta}^{\Delta+W}{d\varepsilon\over
{1+\exp(\omega_{+,\sigma}-\mu)/T}}.
\label{eq:n3}
\end{eqnarray}
The equation for the total number of particles
$2=\sum_{\lambda,\sigma}\langle n_{\lambda,\sigma}\rangle$ together
with Eqs.~(\ref{eq:n2},\ref{eq:n3}) form a closed set of equations for
$\rho$, $m$, and $\mu$, which have to be solved in general
numerically.

At $T=0$, the model can be solved analytically.  For sufficiently
large bandwidth $W$, the following two phases compete: (i)
semiconductor with $\rho=0$, $m=0$ and (ii) ferromagnetic metal with
$\rho=1$, $m=1$. Their energies are $E_{\rm semi}=-2\Delta-2W$ and
$E_{\rm FM}=-\Delta-3W/2-(I+V)/2-\mu_BB$, respectively.  With changing
bandwidth, at $B=0$ there is a level crossing between the
semiconducting phase and the ferromagnetic metal at
$W=W_c(0)=I+V-2\Delta$. For $W>W_c(0)$, the ferromagnetic phase is stable
only in magnetic fields larger than a critical magnetic field
$\mu_B B_c(0) = (W-W_c(0))/2$.

\begin{figure}
\centerline{\psfig{figure=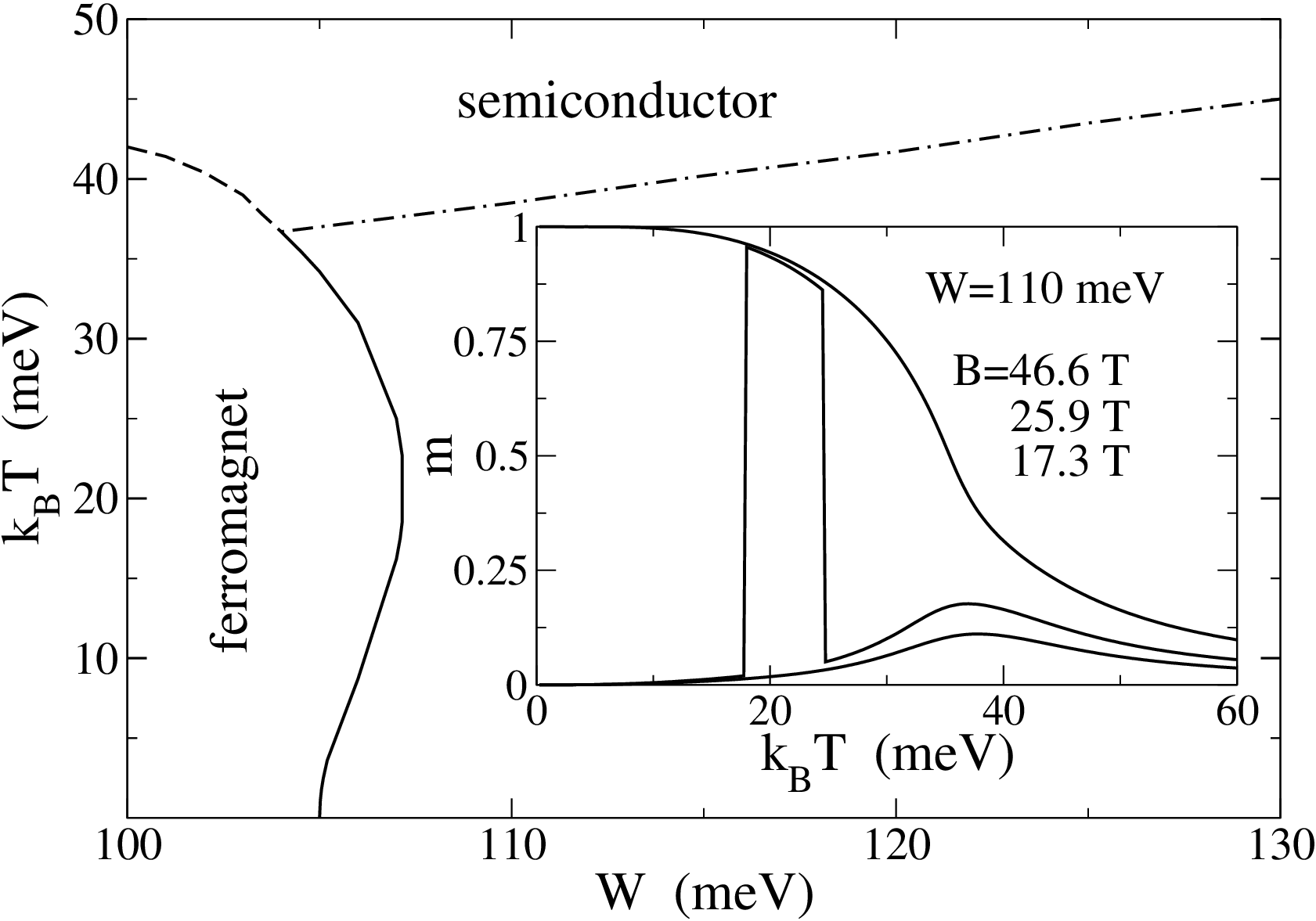,width=7.cm,angle=0}} {\small
FIG. 4: Phase diagram in the $W$-$T$ plane at $B=0$.  The
semiconductor-ferromagnet transition (solid line) is of first order at
low temperatures, up to a critical point at $W_c\approx 104$ meV and
$T_c\approx 37$ meV. At higher temperatures, the transition is of
second order (dashed line). The spin susceptibility in the
paramagnetic phase peaks at temperatures indicated by the dash-dotted
line.  The inset shows the magnetization as a function of temperature
for a nearly critical alloy composition in an applied magnetic field.}
\end{figure}

We take $\Delta=20$ meV in qualitative agreement with experiment and
the effective interaction parameters $I=80$ meV and $V=65$ meV close
to the values proposed in Ref.~\onlinecite{Anisimov96}.  The resulting
phase diagram for $B=0$ shown in Fig.~4 implies that the
phenomenological bandwidth of FeGe, $W_{\rm FeGe}$ is less than
$W_c(0)$ ($W_c(0)$=105~meV).  Note that the Curie temperatures
predicted by our model for $W<W_c(0)$, $T_c\sim 400$ K, are in
reasonable agreement with the experimentally observed transition
temperature to a long-period spiral in FeGe, $T_N\approx 280$
K~\cite{Lebech89}. The phenomenological bandwidth of FeSi, $W_{\rm
FeSi}$,which is greater than $W_c(0)$, can be determined by fitting
the spin susceptibility.  For $W\geq W_c(0)$, the spin susceptibility
develops a strong peak around room temperature, confirming the
interpretation of Anisimov {\it et al.}  The peak position as a
function of $W$ is shown in Fig.~4. Fitting the peak value of $\chi$
to $\approx 27\mu_B^2/{\rm eV}$~\cite{Jaccarino67}, we estimate
$W_{\rm FeSi}\approx 130$~meV.

Nearly critical semiconducting alloys with $W$ slightly in excess of
$W_c(0)$ should exhibit metamagnetic transitions to metallic phases at
experimentally accessible magnetic fields.  For instance in a sample
with $W=110$~meV, the transition occurs at $B_c(0)=43.1$~T at
$T=0$. The temperature evolution of the magnetization $m$ for such a
sample for magnetic fields in the vicinity of the first-order
transition is shown in the inset in Fig.~4. Note that at low
temperatures the critical magnetic field lowers with increasing
temperature. This follows simply from the larger electronic entropy of
the metallic phase.  At higher temperatures the phase boundary in our
model bends towards smaller bandwidths $W$, in qualitative agreement
with the higher susceptibility of small bandwidth systems to various
forms of symmetry breaking transitions. The different low- and
high-$T$ slopes of the phase boundary lead to a reentrant $B=0$ phase
diagram and also to reentrant metamagnetic transitions in a narrow
field range, as shown explicitly in the inset in Fig.~4 for
$B=25.9$~T.

We should like to point out that not only the magnetic properties are
anomalous at the metamagnetic transition. Since the transition is
between a {\it semiconductor} and a ferromagnetic {\it metal}, large
magnetoresistance is to be expected. As a result the critical endpoint
of this metamagnetic transition should show particularily interesting
magnetoresistance behavior at temperatures around room temperature.

\begin{figure}
\centerline{\psfig{figure=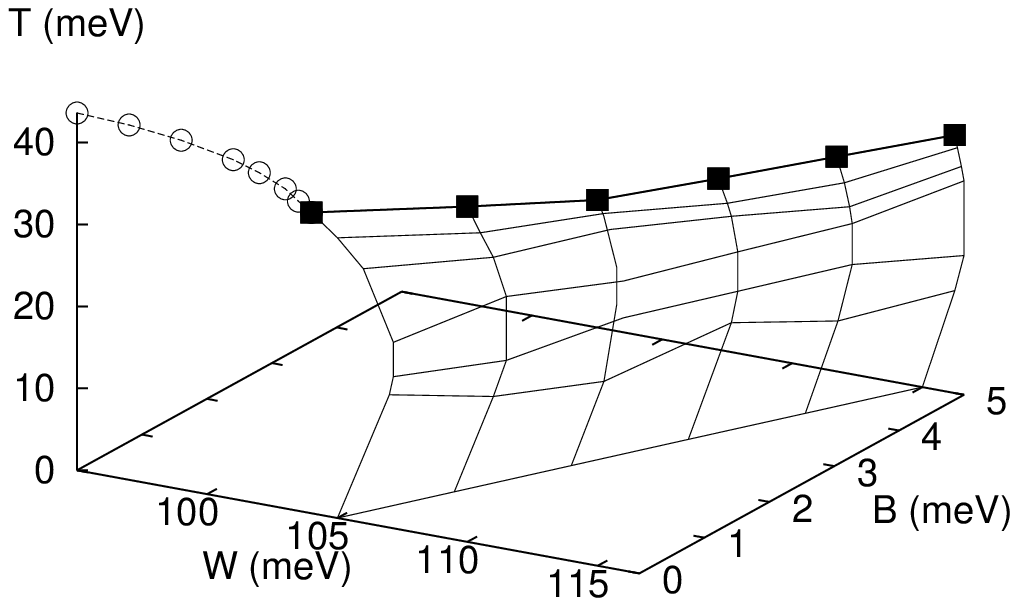,width=9.cm,angle=0}} {\small
FIG. 5: Complete phase diagram of the isoelectronic alloys
FeSi$_{1-x}$Ge$_x$. The semiconducting phase at large bandwidths $W$
(small Ge content $x$) is separated from the ferromagnetic metal phase
at small $W$ (large $x$) by a surface of first-order transitions
terminating in a line of critical points (solid squares) around the
room temperature. At $B=0$ and at Curie temperatures which lie above
the temperature of the critical point, there is a second-order phase
transition between the paramagnetic and ferromagnetic phases (open
circles).}
\end{figure}

In conclusion, we have calculated the complete phase diagram of the
isoelectronic alloys FeSi$_{1-x}$Ge$_x$ with varying $T$ and $B$.  Our
main results are summarized in a three-dimensional phase diagram,
Fig.~5, which shows that, depending on $x$, $T$, and $B$, the alloy
FeSi$_{1-x}$Ge$_x$ can be a semiconductor or a ferromagnetic
metal. The two phases are separated by a surface of first-order
transitions which terminates at high temperatures in a line of
critical points.  Note, this phase diagram resembles that found by
Pfleiderer {\it et al} \cite{MnSi} for isostructural MnSi under
pressure and magnetic field but with the important difference that all
phases of MnSi are metallic so that a semiconductor-metal transistion
is not involved.  We predict that, for intermediate values of the Ge
content $x$, this transition can be realized as a metamagnetic
transition at experimentally accessible magnetic fields.  Large
magnetoresistance is predicted at such a transition.  Another possible
way to realize the transition at experimentally accessible magnetic
fields would be to apply pressure to FeGe.

VIA, VVM, AOS and RH thank the Center for Theoretical Studies, ETH
Zurich and the Swiss National fond for support.  Partial support by
the Slovak Grant Agency VEGA under Grant No. 1/9177/02 and Russian
Foundation For Basic Research grant RFFI-01-02-17063 is also
acknowledged.  We are especially grateful to Zachary Fisk for pointing
out the experimental results on FeGe to us and for communicating his
unpublished results on the FeSi$_{1-x}$Ge$_x$ alloys.

\end{multicols}

\begin{references}
\bibitem{DM} Because of the neglect of spin-orbit coupling in these
calculations, the Dzyaloshinskii-Moria interactions which cause
the long period spiral order in FeGe, are not included and
ferromagnetic order results.
\bibitem{alloy} The electronic structure of FeSi$_{1-x}$Ge$_x$
alloys was calculated in the approximation, where in TBLMTO 
calculation scheme\cite{lmto} the potential parameters for non-metallic
atoms were taken as the corresponding average of the Si and Ge
parameters.
\end{references}
\end{document}